\documentclass[final,authoryear,5p]{elsarticle}

\usepackage{graphicx}
\usepackage{caption,subcaption}

\usepackage[fleqn]{amsmath}

\usepackage{amssymb}

\usepackage[ps2pdf,%
a4paper=true,%
breaklinks=true,%
colorlinks=true,%
pdfauthor={Jovanovic et al.},%
pdftitle={Composite profile of the Fe Kalpha spectral line emitted from a 
binary system of supermassive black holes}%
]{hyperref}

\journal{Advances in Space Research}

\begin{document}

\begin{frontmatter}



\title{Composite profile of the Fe K$\alpha$ spectral line
emitted from a binary system of supermassive black holes\tnoteref{footnote1}}
\tnotetext[footnote1]{This research is part of the project 176003 ''Gravitation 
and the large scale structure of the Universe'' supported by the Ministry of
Education, Science and Technological Development of the Republic of Serbia.
T.B. acknowledges support from the NSF under grant no. AST-12011677. We would 
also like to thank the reviewers for their very helpful comments.}


\author{P. Jovanovi\'{c}\corref{cor}}
\address{Astronomical Observatory, Volgina 7, 11060 Belgrade 38, Serbia}
\cortext[cor]{Corresponding author}
\ead{pjovanovic@aob.rs}

\author{V. Borka Jovanovi\'{c}, D. Borka}
\address{Atomic Physics Laboratory (040), Vin\v{c}a Institute of Nuclear 
Sciences, University of Belgrade, P.O. Box 522, 11001 Belgrade, Serbia}
\ead{vborka@vinca.rs, dusborka@vinca.rs}

\author{T. Bogdanovi\'{c}}
\address{Center for Relativistic Astrophysics, School of Physics, Georgia 
Institute of Technology, 837 State Street, Atlanta, GA 30332-0430, USA}
\ead{tamarab@gatech.edu}

\begin{abstract}

We used a model of a relativistic accretion disk around a supermassive black 
hole (SMBH), based on ray-tracing method in the Kerr metric, to study the 
variations of the composite Fe K$\alpha$ line emitted from two accretion disks 
around SMBHs in a binary system. We assumed that the orbit of such a binary 
is approximately Keplerian, and simulated the composite line shapes for 
different orbital elements, accretion disk parameters and mass ratios of the 
components. The obtained results show that, if observed in the spectra of some 
SMBH binaries during their different orbital phases, such composite Fe 
K$\alpha$ line profiles could be used to constrain the orbits and several 
properties of such SMBH binaries.

\end{abstract}

\begin{keyword}

black holes: black-hole binaries; galaxies: active; accretion and accretion 
disks; spectral lines; X-ray emission spectra

\end{keyword}

\end{frontmatter}

\parindent=0.5 cm

\section{Introduction}

Binary systems of supermassive black holes (SMBHs) originate in galactic 
mergers, and it is believed that their coalescences represent the most powerful 
sources of low-frequency gravitational-waves (from 0.1 to 100 mHz) which
could provide a significant amount of information not only about the SMBH 
binaries themselves \citep[see e.g.][]{dep03}, but also about the early 
universe, the structure and nature of spacetime, and formation and structure of 
galaxies. They are expected to be detected in near future by space-based 
interferometers, such as \emph{New Gravitational-Wave Observatory (NGO)}, also 
known as \emph{evolved Laser Interferometer Space Antenna (eLISA)} \citep[see 
e.g.][]{ama13}, as well as by ASTROD I \citep{bra12}. These measurements of 
gravitational radiation will be used for testing General Relativity with higher 
sensitivity, improving our understanding of gravity, and will complement 
traditional astronomical observations based on the electromagnetic spectrum.

On the other hand, electromagnetic radiation in different spectral bands 
emitted during such coalescences of SMBHs represents the most direct evidence 
for the formation of their binary systems, as well as their essential 
observational signatures \citep[see e.g.][]{bog09a}. At some stage during 
galactic merger, two SMBHs initially carried within the bulges of their host 
galaxies, will become gravitationally bound and will start to orbit around their 
center of mass with velocities of a few thousand km~s$^{-1}$. In such cases, 
accretion of the surrounding matter on both SMBHs could be expected, and as a 
result, a strong X-ray emission in the broad Fe K$\alpha$ line at 6.4 keV might 
be observed.

It was first showed by \citet{fab89} that the broad Fe K$\alpha$ 
emission line could be well modeled by fluorescent emission from the inner 
parts of a relativistic accretion disk around a SMBH, and that these regions of 
the disk could be mapped using the shape and variability of the line, since both 
of them are affected by Doppler and gravitational redshifts produced in vicinity 
of the central SMBH \citep[for more details see e.g. reviews 
by][]{rey03,fab10}. The first convincing observational proof for the existence 
of the relativistically broadened Fe K$\alpha$ line was found by \citet{tan95} 
in the X-ray spectra of Seyfert 1 galaxy MCG-6-30-15, obtained by Japanese 
\emph{ASCA} satellite.

The disks in binaries with arbitrary eccentricity and mass 
ratio are truncated due to ordinary and eccentric Lindblad resonances, and 
their sizes depend on binary mass ratio and eccentricity \citep[see e.g.][and 
references therein]{lin79a,lin79b,art94,hay11}. Therefore, stability of the 
orbits of the SMBH binaries could limit the sizes of the accretion disks around 
the components, as well as the size of their circumbinary accretion disk.
Thus, the Fe K$\alpha$ line emitting regions could have different structures, 
depending on the mass ratios of such binaries, properties of their accretion 
disks and separation between their components. In some cases, the secondary SMBH 
could be embedded in the accretion disk around the primary SMBH causing an empty 
annular gap in it which, as a consequence, could produce imprints in form of 
ripples in the observed Fe K$\alpha$ line profile \citep[see e.g.][]{mck13}. In 
other cases, the binary clears a low density ''hole'' or cavity in the center of 
a circumbinary disk \citep[see e.g.][and references 
therein]{bog09a,bog09b,bog11}, and the Fe K$\alpha$ line emission 
arising from the accretion disks around primary and secondary SMBHs is then
affected by the Doppler shifts due to the orbital motion of the binary. 
\citet{ses12} have shown that such double Fe K$\alpha$ lines could be
in principle identified and used to estimate the properties of the SMBHs, given 
the high spectral resolution observations with a next generation X-ray 
observatory. Such SMBH binaries are of special significance because a 
technique for their detection which utilizes the Doppler shifts in their 
spectra, although not unique, is one of the simplest and the most 
straightforward \citep[see e.g.][]{bog09a}. In further text we will refer to the 
total Fe K$\alpha$ line emission from both primary and secondary disks of such a 
system as ''composite'', and to the emission from each of individual disks as 
''constituent''.

In our previous investigations we developed numerical simulations of 
X-ray emission from a relativistic accretion disk around a SMBH, based on 
ray-tracing method in Kerr metric, and successfully applied these simulations 
for studying: (i) properties of SMBHs in some observed Active Galactic Nuclei 
(AGN) from the shapes of their Fe K$\alpha$ lines \citep{jov11}, (ii) 
observational effects of strong gravity in vicinity of SMBHs \citep{jov08a}, 
(iii) photocentric variability of QSOs, as well as variability of their X-ray 
and optical lines, due to perturbations in their accretion disks 
\citep{jov10,pop12a}, (iv) influence of gravitational 
microlensing on radiation from X-ray, UV and optical emitting regions of 
accretion disks of AGN \citep{jov08b,jov09a,pop03a,pop03b,pop05,pop06}. For 
reviews of our previous investigations of influence of single and binary SMBHs 
in AGN on emission lines see e.g. \citet{jov12,jov09b} and \citet{pop12b}.

Here we use these simulations for studying the composite Fe K$\alpha$ line 
profiles emitted from both components of a SMBH binary system. Usually the 
broad optical lines, like H$\alpha$ and H$\beta$, are used for studying the 
SMBH binaries \citep[see e.g.][]{bog08,bog09b,bon12,pop12b}. However, it is 
very difficult to identify the components in their profiles which vary due to 
Doppler shift. Such components are usually in the form of broad and faint 
''bumps'' embedded in the line profile and their variability usually invokes a 
non-binary interpretation, since it has been seen in long-term monitoring of 
disk emitters, while at the same time, a SMBH binary interpretation is still 
considered just as a possible alternative \citep[see e.g.][and references 
therein]{she13}. In contrast, the Fe K$\alpha$ line has more asymmetric profile 
with much sharper peaks, and in principle, it should be easier to identify its 
constituent and varying components. Here we study whether this property of the 
composite Fe K$\alpha$ line could be exploited for studying the SMBH binaries.

The paper is organized as follows: in the second section we present the models 
and parameters of a binary system of SMBHs and relativistic accretion disks 
around its components, in the third section we describe our simulations and the 
resulting composite profiles of the Fe K$\alpha$ spectral line for several 
cases, and finally in the fourth section we outline our conclusions.

\begin{figure*}[ht!]
\begin{center}
\includegraphics*[width=\textwidth]{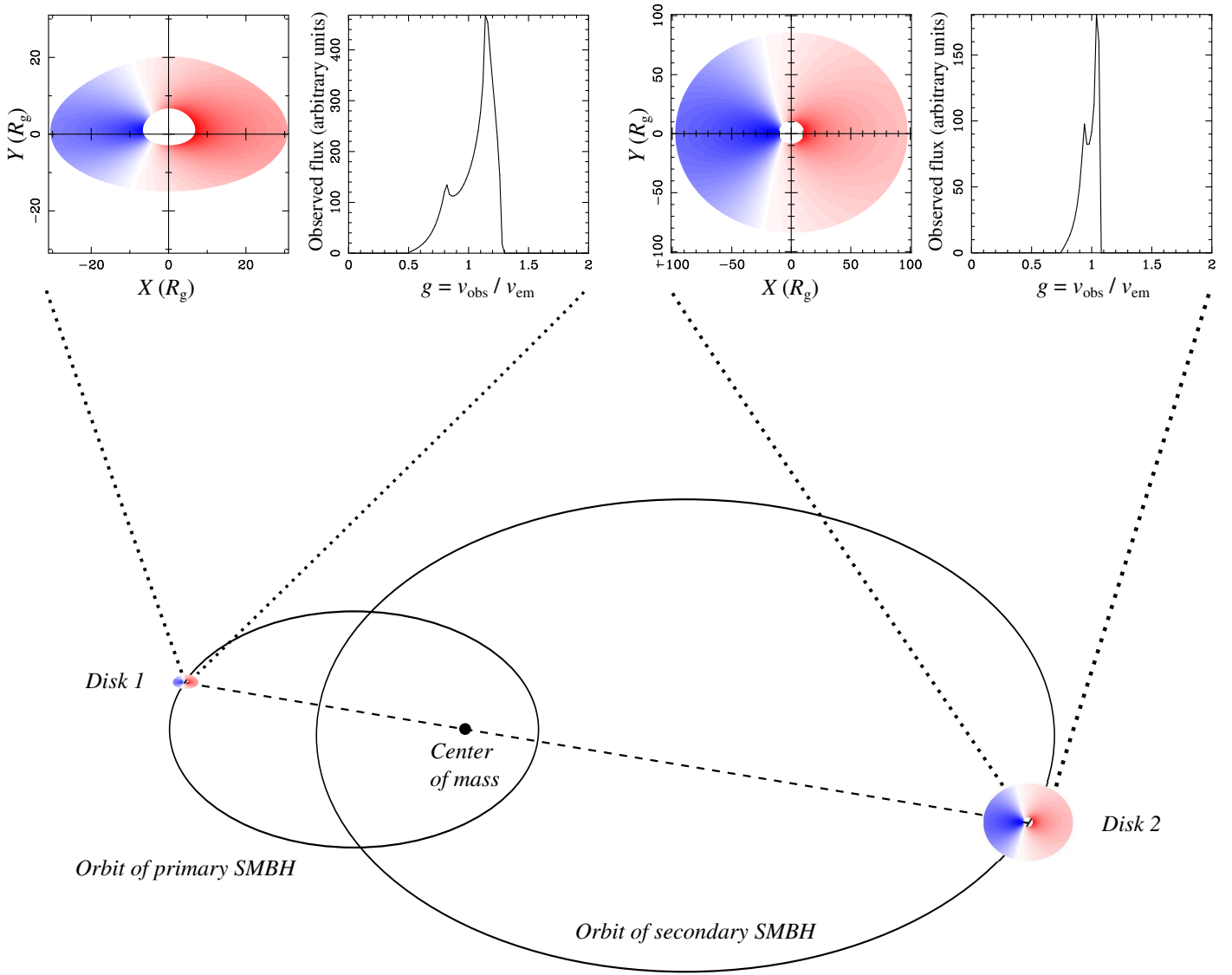}
\end{center}
\caption{An illustration of two accretion disks around the components of a 
binary system of SMBHs, rotating along a Keplerian orbit. \textit{Disk 1} and 
\textit{disk 2}, as well as the corresponding simulated profiles of the Fe 
K$\alpha$ spectral line emitted from them, are zoomed in the top part of the 
figure. See Table \ref{tab1} for the disk parameters.}
\label{fig1}
\end{figure*}

\section{Iron K$\alpha$ emission from SMBH binaries}

\subsection{Model of relativistic accretion disk around a SMBH}

We modeled the emission from accretion disk by numerical
simulations based on ray-tracing method in Kerr metric, taking into account only
photon trajectories reaching the observer's sky plane. For more details about 
this method see e.g. \citet{fan97} and \citet{cad98} and references therein, 
but briefly, in this method one divides the image of the disk on the observer's 
sky into a number of small elements (pixels) and for all of them the photon
trajectories are traced backward from the observer until their intersection 
with the plane of the disk, by pseudo-analytic integration of the corresponding 
geodesics in Kerr space-time. For each photon the ratio between the observed 
$\nu_{obs}$ and emitted $\nu_{em}$ frequencies (i.e. the redshift factor $g$) 
is calculated:
\begin{equation}
\label{shift}
g=\frac{\nu _{obs}}{\nu _{em}}=\frac{1}{1+z}.
\end{equation}
A simulated image of an accretion disk, as would be seen by a distant observer 
by a high resolution telescope, is obtained by coloring all pixels according to 
the corresponding values of $g$. The corresponding  simulated line profile $F 
\left(g\right)$ is obtained from the total observed flux distribution at the 
observed energy $E_{obs}$ \citep{fan97}: 
\begin{equation}
\label{flux}
F \left( {E_{obs}}  \right) = {\int\limits_{image} {\varepsilon
\left({r} \right)}} g^{4}\delta \left( {E_{obs} - gE_{0}}  \right)d\Xi ,
\end{equation}
where $\varepsilon \left( {r} \right) = \frac{{\varepsilon}_0}{4\pi}r^{-p}$ is 
the surface emissivity of the disk which varies with radius as a power law with 
emissivity index $p$, $d\Xi$ is the solid angle subtended by the disk in the 
observer's sky and $E_{0}$ is the rest energy of the line.

Here we use this model of accretion disk to study the composite Fe K$\alpha$ 
line profiles emitted from SMBH binaries, taking into account Doppler shifts
during their different orbital phases. For that purpose, and in order to study 
some more realistic cases, we modeled two disks (denoted as \textit{disk 1} and 
\textit{disk 2}) with different radii and inclinations, because such disks emit 
significantly different profiles of the Fe K$\alpha$ line (as it can be seen 
from the top part of Fig. \ref{fig1}). Both disks surround the slowly 
rotating Kerr SMBHs, having the same spin (i.e. angular momentum $J$ per unit 
mass $M$ of the SMBH) of $J/M c=0.1$. The first disk is extending from $R_{ms}$ 
to $30\ R_g$ (where $R_{ms}$ is radius of the marginally stable orbit, 
$R_g=GM/c^2$ is the gravitational radius, and $G$ and $c$ are well 
known constants) and has inclination of $60^\circ$ (\textit{disk 1}). The second 
one is extending from $10\ R_g$ to $100\ R_g$ and has inclination of $30^\circ$ 
(\textit{disk 2}). We assumed that the inclinations of the disks do not change 
with the time, or along the orbit, as well as that both of them have the same 
emissivity index of $p=2$ \citep[see e.g.][and references therein]{jov09b}. 
Also, since the Fe K$\alpha$ line originates from the innermost parts of an 
accretion disk close to the marginally stable orbit around a SMBH \citep[see 
e.g.][and references therein]{jov12}, a circumbinary disk is assumed not to 
contribute to the total Fe K$\alpha$ line emission of the system. 

\subsection{Keplerian radial velocity curves of SMBH binaries}

It is well known from the theory of close binary stars that the radial 
velocities $V_{1,2}^{rad}$ of a binary system can be calculated 
from the following orbital elements: eccentricity $e$, inclination $i$, 
longitude of pericenter $\omega$, separation between the components $a$ and 
their masses $M_{1,2}$, according to the following expression 
\citep[see e.g.][]{hild01}:
\begin{equation}
\label{eq_vrad}
V_{1,2}^{rad}\left(\theta\right) = {K_{1,2}}\left[ {\cos \left( 
{\theta  + 
\omega } \right) + e \cdot \cos \omega } \right] + \gamma .
\end{equation}
In previous equation $\theta$ is true anomaly, $\gamma$ is systemic velocity 
and $K_{1,2}$ are semiamplitudes of the velocity curves:
\begin{equation}
\label{eq_k12}
K_{1,2} = \frac{{2\pi {a_{1,2}}\sin i}}{{P\sqrt {1 - {e^2}} }},
\end{equation}
where the semimajor axes $a_{1,2}$ are found from: $a = {a_1} + {a_2}$ and 
${M_1}{a_1} = {M_2}{a_2}$. Radial velocity curves as functions of time can be 
calculated from (\ref{eq_vrad}) using Kepler's equation, and orbital period of 
the binary system can be determined from the third Kepler's law:
\begin{equation}
\label{period}
P^2 = \frac{{4{\pi ^2}{a^3}}}{{G\left( {{M_1} + {M_2}} \right)}}.
\end{equation}

An illustration of a Keplerian orbit of a SMBH binary is shown in bottom part 
of Fig. {\ref{fig1}. In all our simulations, we assumed that the masses of 
primary and secondary SMBHs are ${M_1} = 1 \times {10^8}{M_ \odot}$ and ${M_2} 
= q\cdot M_1 $, where the mass ratio $q$ is taken to be 1 or 0.5 \citep[see 
e.g.][]{bon12}. In order to avoid and neglect the self-lensing effects between 
the components in the binary system \citep[see e.g.][]{gou95}, we studied only 
significantly inclined orbits with $i = 30^\circ$ and $i = 60^\circ$, and with 
relatively large separation between the components of $a = 0.01$ pc.
Therefore, we modeled two binary systems of SMBHs (denoted as \textit{binary 1} 
and \textit{binary 2}) with different mass ratios, orbital inclinations and 
eccentricities. In the case of the first one: $q = 1$, $i = 60^\circ$, $e = 
0.75$ (\textit{binary 1}), while in the case of the second one: $q = 0.5$, $i = 
30^\circ$, $e = 0$ (\textit{binary 2}). In both cases $\omega = 90^\circ$ and 
systemic velocity $\gamma  = 0$. The corresponding Keplerian radial velocity 
curves are presented in Fig. \ref{fig2}.

\begin{figure}[ht!]
\begin{center}
\includegraphics*[width=\columnwidth]{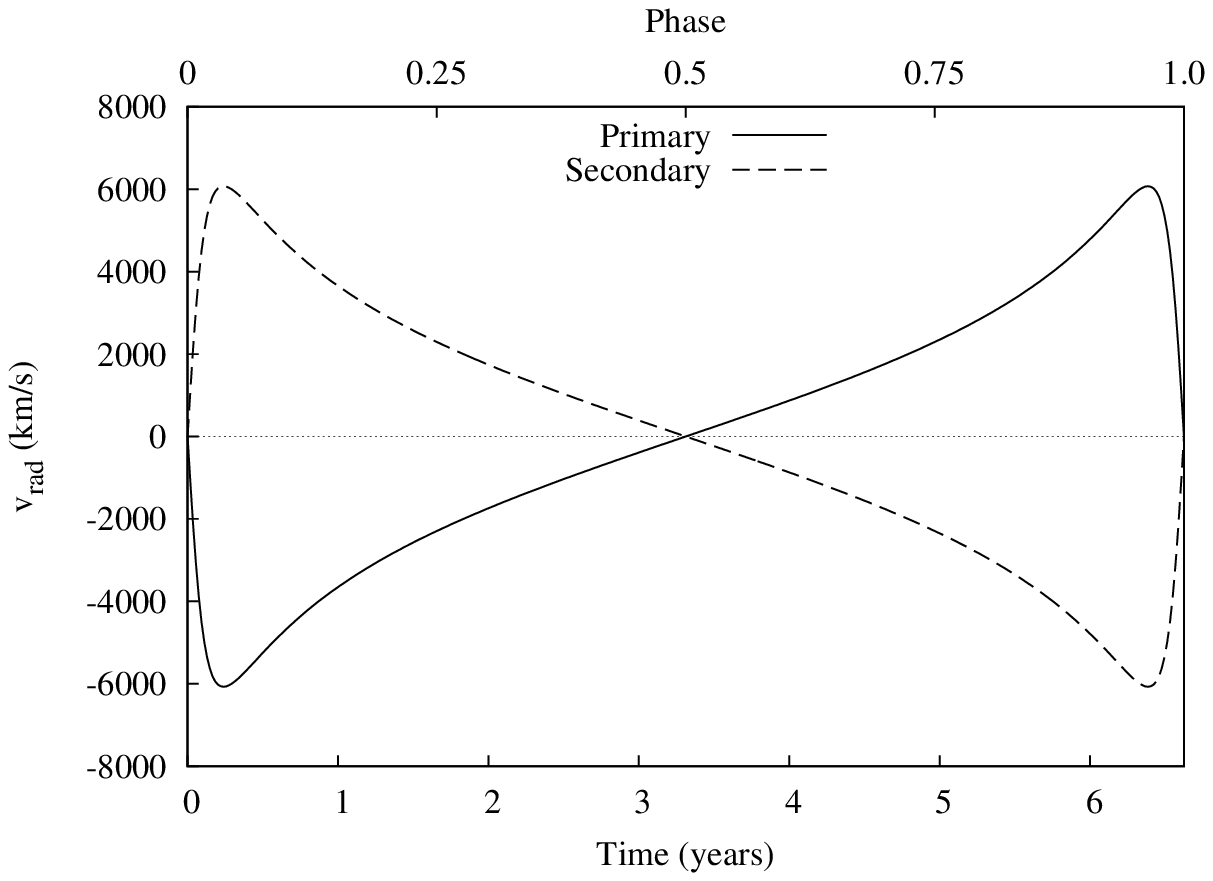} \\
\vspace*{0.5cm}
\includegraphics*[width=\columnwidth]{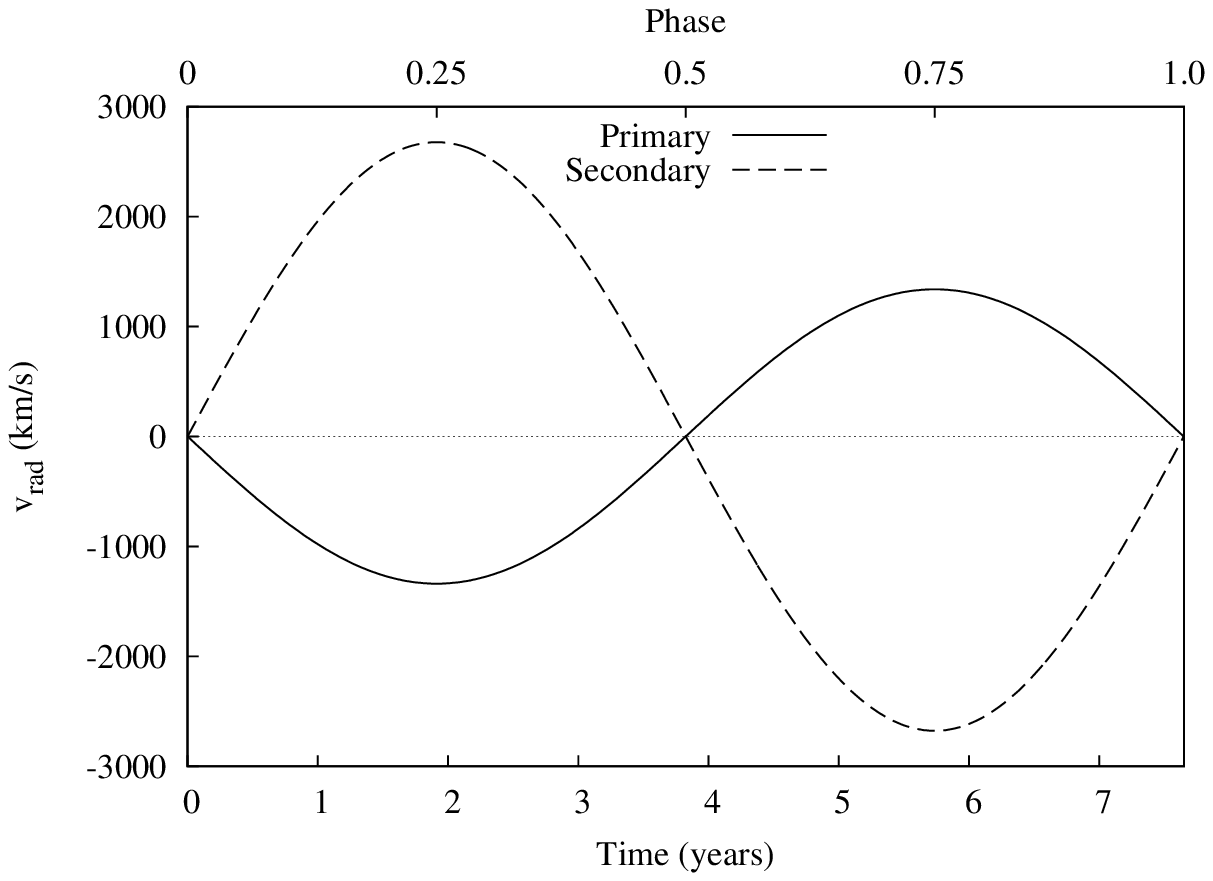}
\end{center}
\caption{Keplerian radial velocity curves for two binary systems of SMBHs: 
\textit{binary 1} (top) and \textit{binary 2} (bottom). See Table \ref{tab2} 
for the parameters of these systems.}
\label{fig2}
\end{figure}

\subsection{Composite profiles of the Fe K$\alpha$ line}

Taking into account that for Keplerian orbits $V_{1,2}^{rad} \ll c$, the 
corresponding contributions of Doppler effect to the redshift $z$ in expression 
(\ref{shift}) are $\approx V_{1,2}^{rad} / c$.  Hence, a composite profile 
$F\left(g\right)$ of the Fe K$\alpha$ line emitted from both accretion disks 
at some orbital phase, can be calculated from two constituent unshifted 
profiles $F_1 \left(g\right)$ and $F_2 \left(g\right)$ according to:
\begin{equation}
\label{lprofile}
F\left(g\right)=F_1\left(\left[\frac{1}{g}-\frac{V_1^{rad}}{c}\right]^{-1}
\right)+F_2\left(\left[\frac{1}{g}-\frac{V_2^{rad}}{c}\right]^{-1}\right).
\end{equation}

\section{Results and discussion}

We first simulated the constituent Fe K$\alpha$ lines emitted by 
\textit{disk 1} and \textit{disk 2} (see top part of Fig. \ref{fig1}). The
parameters used in these simulations are summarized in Table \ref{tab1}.
In order to quantify the widths of the obtained constituent lines, we measured 
their full width at half maximum (FWHM). In the case of \textit{disk 1} we 
obtained: FWHM$_1$ $\approx 36000$ km~s$^{-1}$ and in the case of \textit{disk 
2}: FWHM$_2$ $\approx 20000$ km~s$^{-1}$, both being in the range of the 
corresponding values typically observed in AGN \citep[see e.g.][]{nan97,nan07}. 

\begin{table}[h!]
\centering
\caption{The parameters of two simulated accretion disks: $J/M c$ - spin of 
the central SMBH, $\theta_{obs}$ - disk inclination, $R_{in}$ - inner radius of 
the disk, $R_{out}$ - outer radius of the disk, $p$ - power law emissivity 
index.}
\small
\begin{tabular}{|c||c|c|c|c|c|}
\hline
& $J/M c$ & $\theta_{obs}\ (^\circ )$ & $R_{in}\ (R_g)$ & $R_{out}\ (R_g)$ & 
$p$ \\
\hline
\hline
\textit{disk 1} & 0.1 & 60 & $R_{ms} = 5.67$ & 30 & 2 \\
\hline
\textit{disk 2} & 0.1 & 30 & 10 & 100 & 2 \\
\hline
\end{tabular}
\normalsize
\label{tab1}
\end{table}

\begin{table}[h!]
\centering
\caption{Mass ratios and orbital elements of two simulated SMBH binaries: $q$ - 
mass ratio (for ${M_1} = 1 \times {10^8}{M_ \odot}$), $a$ - separation between 
the components, $i$ - inclination, $e$ - eccentricity, $\omega$ - longitude of 
pericenter and $\gamma$ - systemic velocity.}
\small
\begin{tabular}{|c||c|c|c|c|c|c|}
\hline
& $q$ & $a$ (pc) & $i\ (^\circ )$ & $e$ & $\omega\ (^\circ )$ & $\gamma$ 
(km/s) \\
\hline
\hline
\textit{binary 1} & 1 & 0.01 & 60 & 0.75 & 90 & 0 \\
\hline
\textit{binary 2} & 0.5 & 0.01 & 30 & 0 & 90 & 0 \\
\hline
\end{tabular}
\normalsize
\label{tab2}
\end{table}

\begin{figure*}[t!]
\begin{subfigure}[b]{\columnwidth}
\centering
\caption{\textit{disk 1} (primary) and \textit{disk 1} (secondary)}
\includegraphics*[width=\columnwidth]{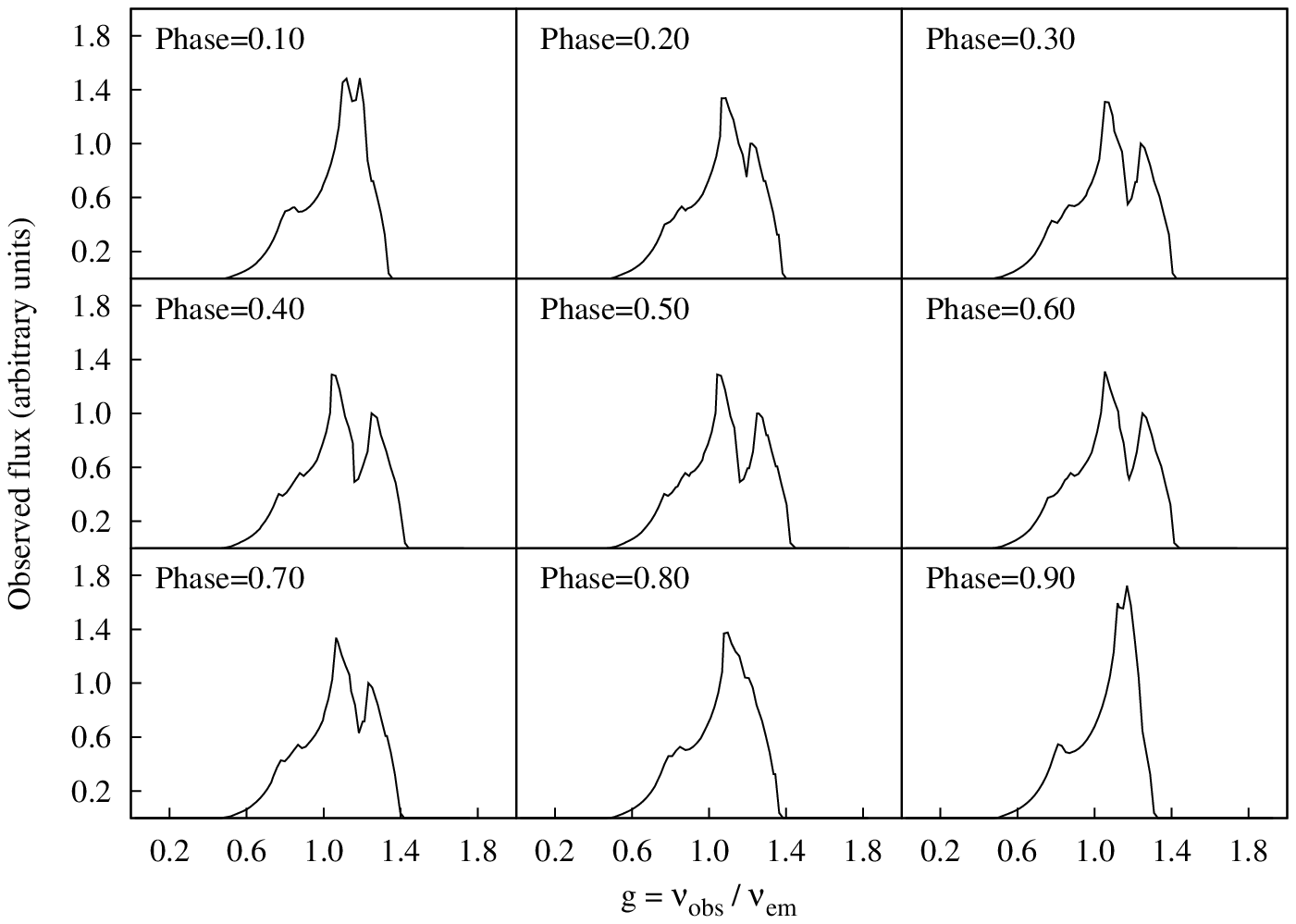}
\label{fig3a}
\end{subfigure}
\hfill
\begin{subfigure}[b]{\columnwidth}
\centering
\caption{\textit{disk 2} (primary) and \textit{disk 2} (secondary)}
\includegraphics*[width=\columnwidth]{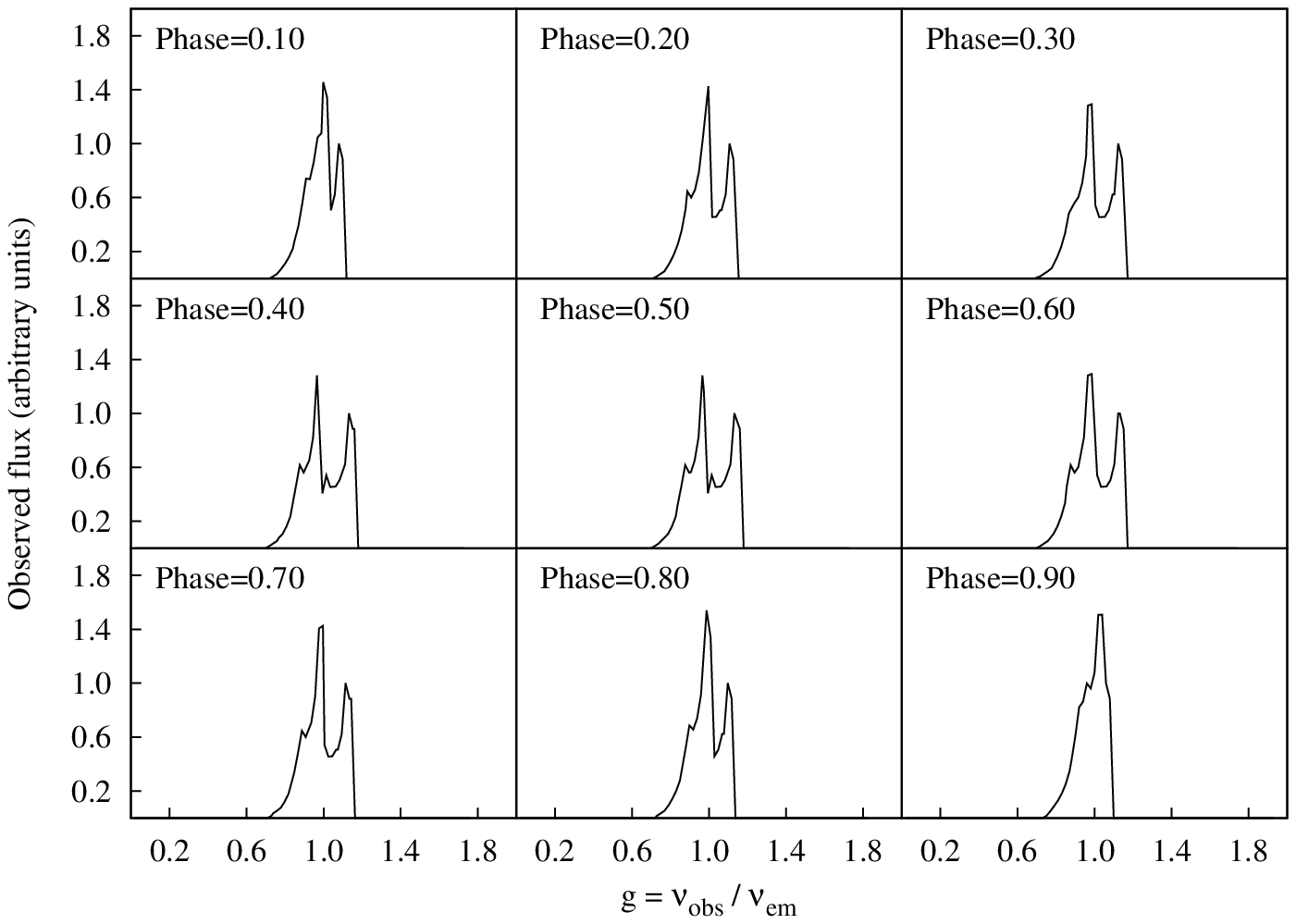}
\label{fig3b}
\end{subfigure}
\\
\begin{subfigure}[b]{\columnwidth}
\centering
\caption{\textit{disk 1} (primary) and \textit{disk 2} (secondary)}
\includegraphics*[width=\columnwidth]{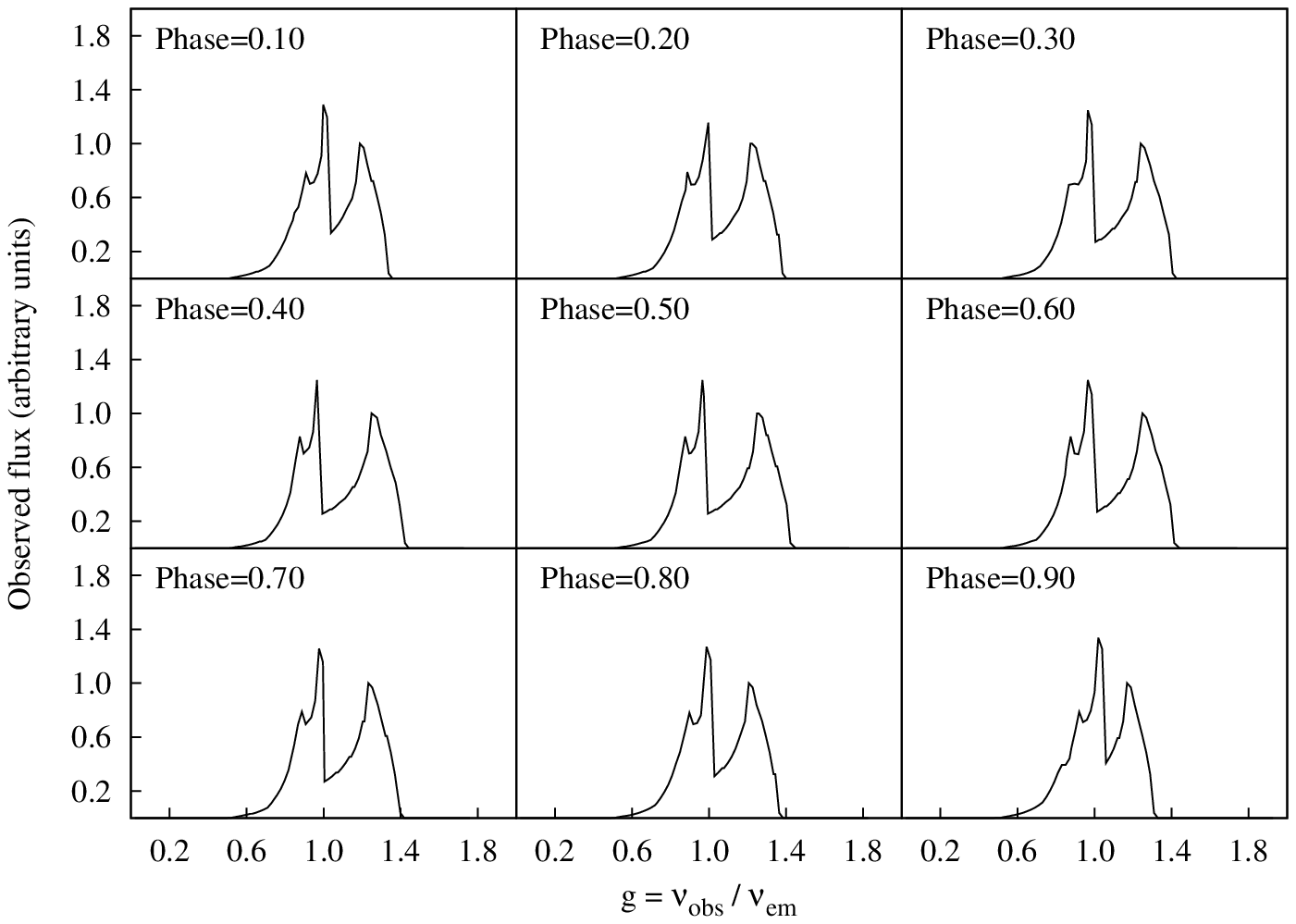}
\label{fig3c}
\end{subfigure}
\hfill
\begin{subfigure}[b]{\columnwidth}
\centering
\caption{\textit{disk 2} (primary) and \textit{disk 1} (secondary)}
\includegraphics*[width=\columnwidth]{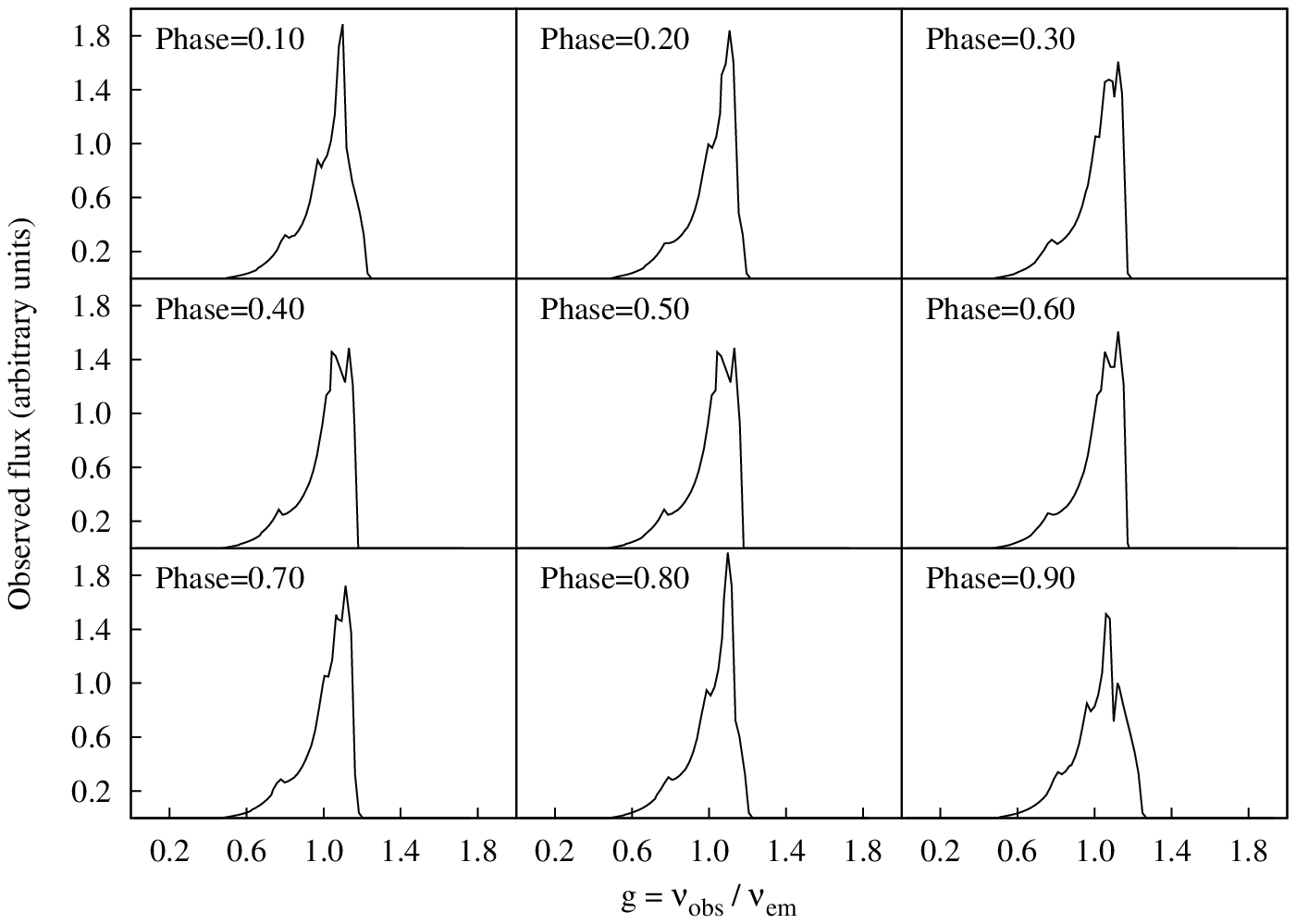}
\label{fig3d}
\end{subfigure}
\caption{Composite profiles of the Fe K$\alpha$ line emitted during different 
orbital phases of the \textit{binary 1}. Model of accretion disk around each 
component is denoted in the caption of each sub panel.}
\label{fig3}
\end{figure*}

\begin{figure*}[t!]
\begin{subfigure}[b]{\columnwidth}
\centering
\caption{\textit{disk 1} (primary) and \textit{disk 1} (secondary)}
\includegraphics*[width=\columnwidth]{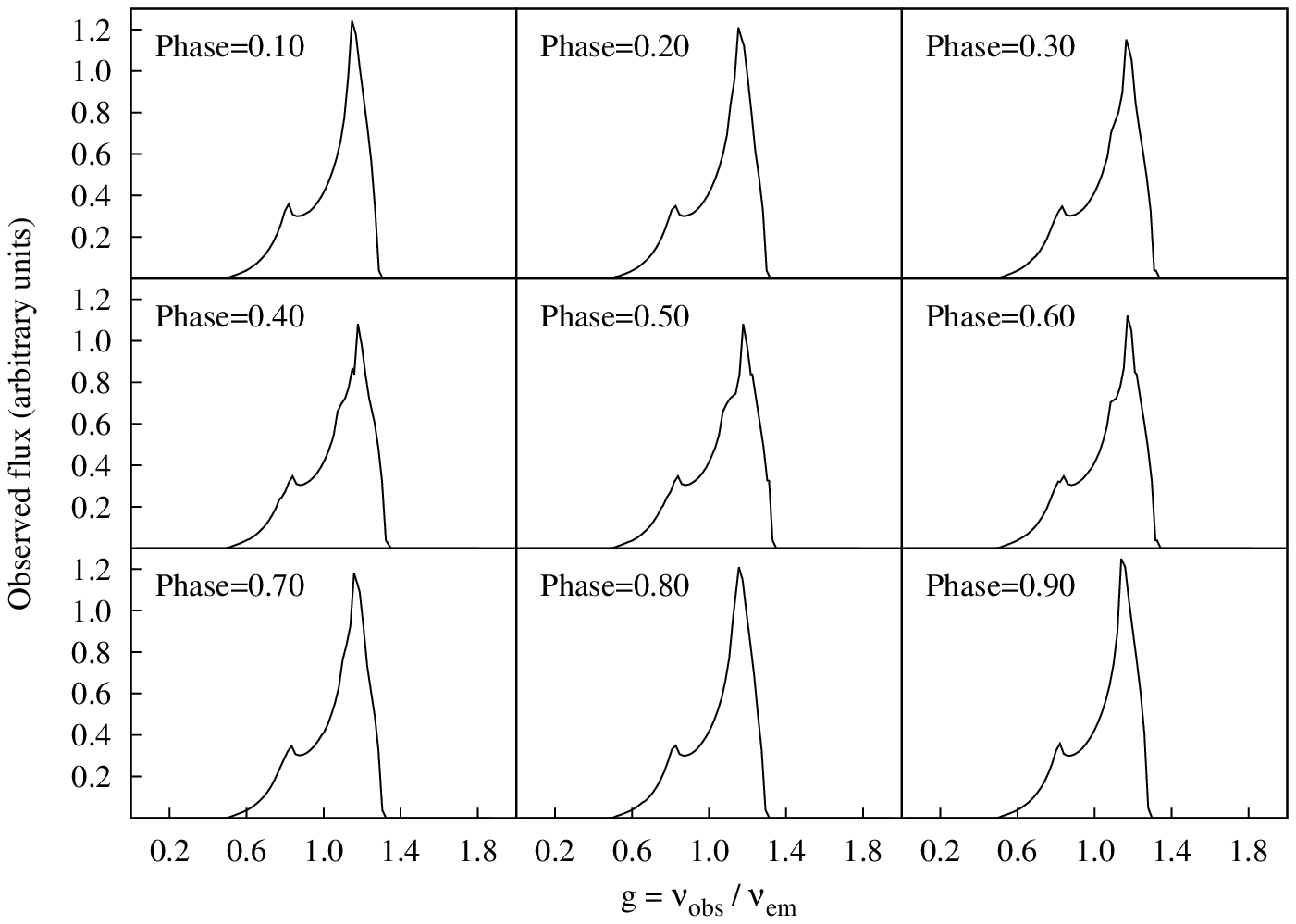}
\label{fig4a}
\end{subfigure}
\hfill
\begin{subfigure}[b]{\columnwidth}
\centering
\caption{\textit{disk 2} (primary) and \textit{disk 2} (secondary)}
\includegraphics*[width=\columnwidth]{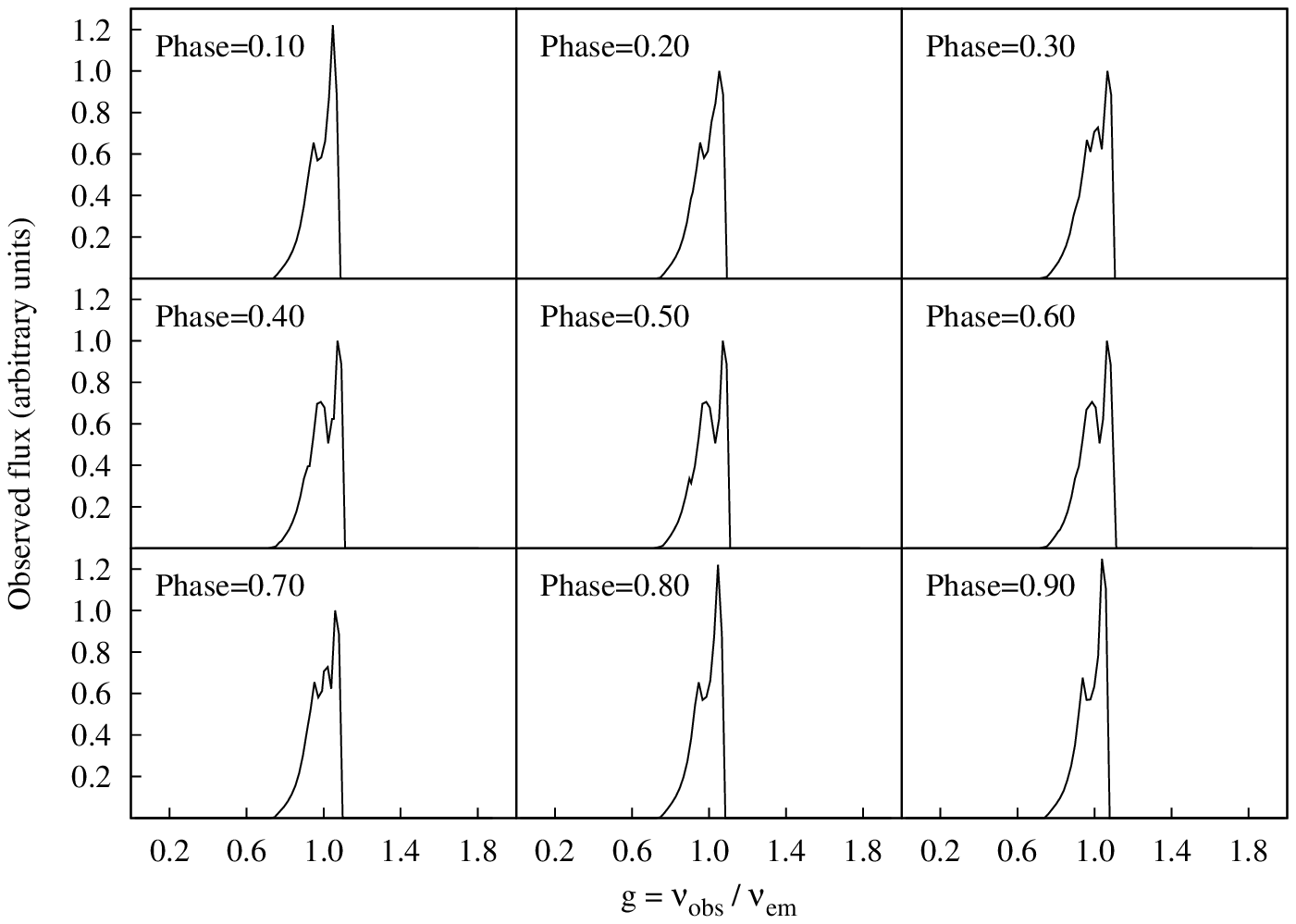}
\label{fig4b}
\end{subfigure}
\\
\begin{subfigure}[b]{\columnwidth}
\centering
\caption{\textit{disk 1} (primary) and \textit{disk 2} (secondary)}
\includegraphics*[width=\columnwidth]{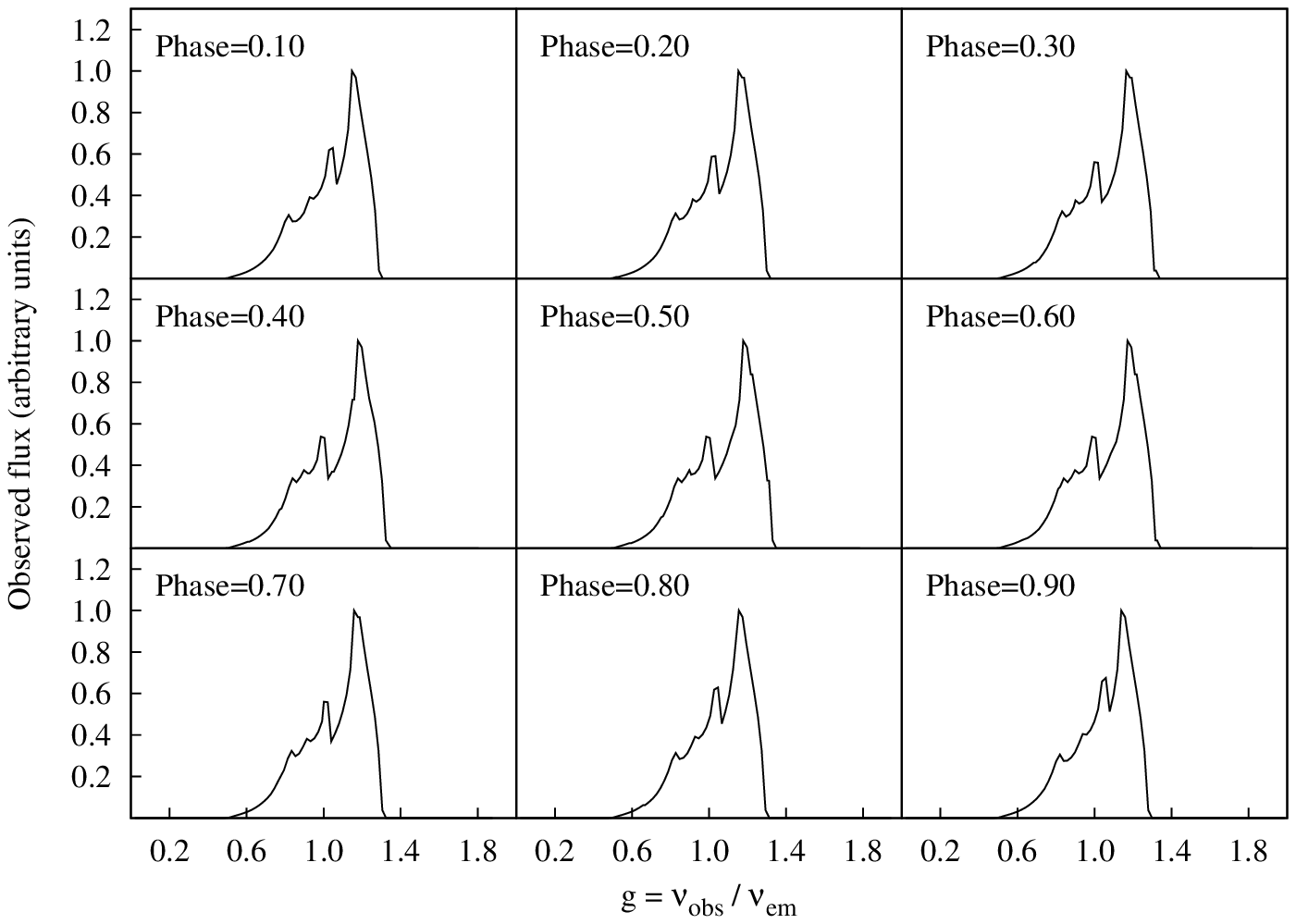}
\label{fig4c}
\end{subfigure}
\hfill
\begin{subfigure}[b]{\columnwidth}
\centering
\caption{\textit{disk 2} (primary) and \textit{disk 1} (secondary)}
\includegraphics*[width=\columnwidth]{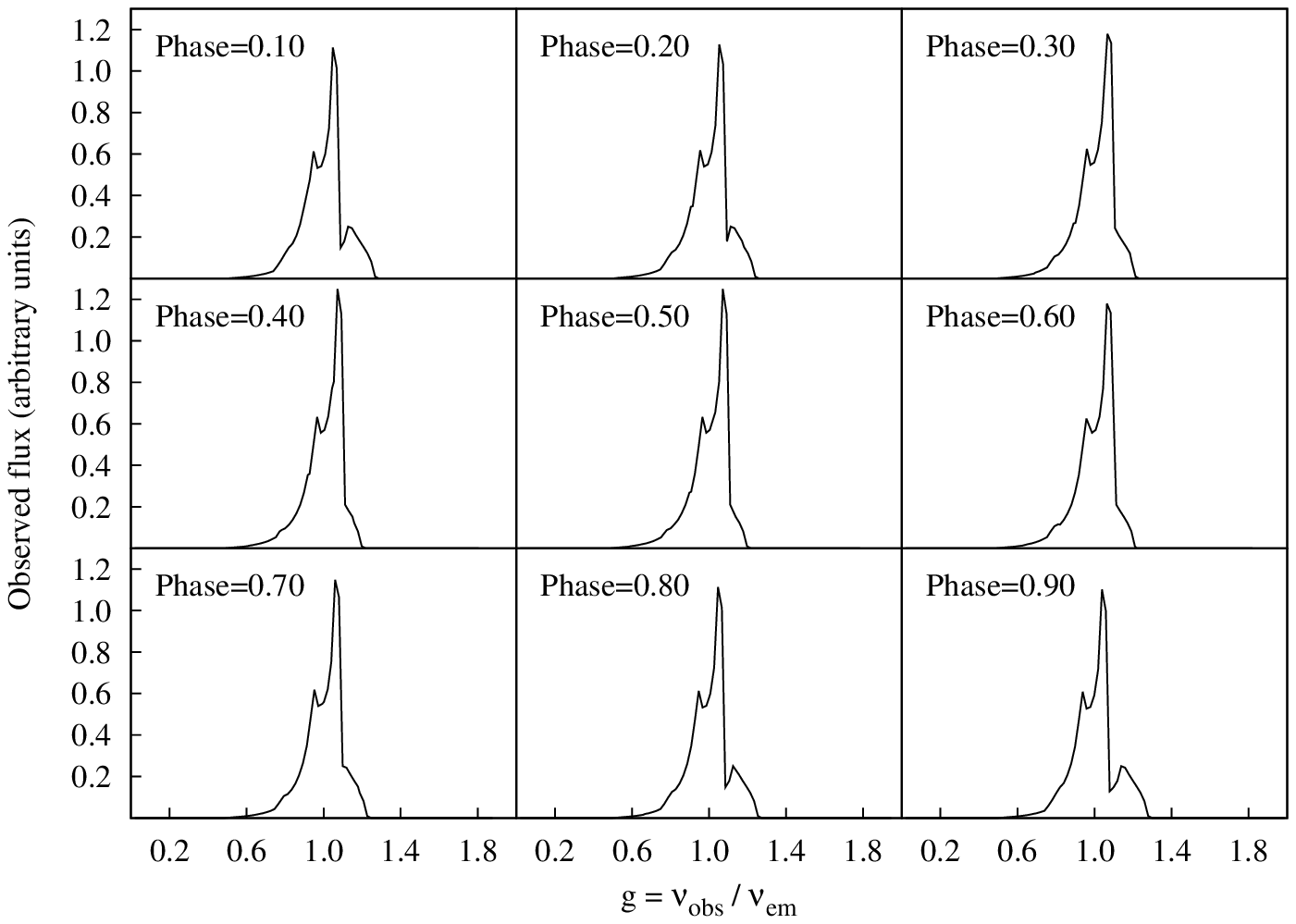}
\label{fig4d}
\end{subfigure}
\caption{The same as in Fig. \ref{fig3}, but in the case of 
the \textit{binary 2}.}
\label{fig4}
\end{figure*}

In the next step, we simulated radial velocity curves of \textit{binary 1} and 
\textit{binary 2} (see Fig. \ref{fig2}), using the parameters presented in 
Table \ref{tab2}. As it can be seen from Fig. \ref{fig2}, both SMBH 
binaries have orbital periods of approximately 7 years and velocity 
semiamplitudes of several thousand km~s$^{-1}$.  In the case of \textit{binary 
1}, the radial velocities of both components reach $\approx \pm 6000$ 
km~s$^{-1}$ at certain orbital phases, which is approximately $17\%$ of 
FWHM$_1$ and $30\%$ of FWHM$_2$. Hence, the Fe K$\alpha$ line emitted from such 
a binary could be significantly affected by the Doppler effect due to orbital 
motion. The smaller Doppler shifts are expected in the case of \textit{binary 
2}, since the radial velocity of its primary reach $\approx \pm 1300$ 
km~s$^{-1}$ and of the secondary $\approx \pm 2700$ km~s$^{-1}$, which is 
$\lesssim 10\%$ of FWHM$_1$ and FWHM$_2$. 

For both SMBH binaries we then simulated the composite Fe K$\alpha$ line 
profiles in the following four cases for accretion disks around the 
primary and secondary components:
\begin{enumerate}
\item \textit{disk 1} (primary) and \textit{disk 1} (secondary); 
\item \textit{disk 2} (primary) and \textit{disk 2} (secondary);
\item \textit{disk 1} (primary) and \textit{disk 2} (secondary); 
\item \textit{disk 2} (primary) and \textit{disk 1} (secondary).
\end{enumerate}
The obtained composite profiles during nine different orbital phases are 
presented in Figs. \ref{fig3} and \ref{fig4}. Due to higher radial 
velocities, the larger Doppler shifts are expected in the case of 
\textit{binary 1}, which can be seen by comparing the corresponding panels of 
Figs. \ref{fig3} and \ref{fig4}. As it can be also seen from Figs. \ref{fig3} 
and \ref{fig4}, these Doppler shifts cause ripple effects and clefts in the 
composite Fe K$\alpha$ line profiles which oscillate with orbital period.

In the case of \textit{binary 1} these effects are very strong, mostly due to 
great mass ratio $q$, and sometimes cause so deep clefts in the composite line 
profiles that the corresponding constituent profiles can be almost completely 
resolved. This is especially the case when the secondary SMBH is surrounded by 
\textit{disk 2}, which emits narrower line (see the panels (b) and (c) of Fig. 
\ref{fig3}), and to a lesser extent when the secondary is accreting through 
\textit{disk 1}, which emits broader line (see the panels (a) and (d) of the 
same figure). This difference indicates that, besides the mass ratio $q$, 
the disk inclination also has a significant impact on the composite 
profile variability, since a more inclined disk emits a broader line (see 
top part of Fig. \ref{fig1}). The similar result was obtained by \citet{yul01},
who also studied a possibility to probe the SMBH binaries by their observed Fe 
K$\alpha$ line profiles. However, in contrast to \citet{yul01}, our results 
demonstrate that the orbital motion of the SMBH components should not be 
neglected, since it could cause detectable Doppler shifts in the constituent Fe 
K$\alpha$ line profiles.

If such composite profiles of the Fe K$\alpha$ line with resolvable constituent 
components were observed, it would be possible to measure their Doppler 
shifts, which could be then used for reconstructing the radial velocity curves 
of the SMBH binary \citep[as shown in][]{bon12}. One could then fit Keplerian 
orbits to these velocity curves in order to determine the orbital elements and 
mass ratio of the components. Providing an independent estimate for orbital 
inclination, the masses of both components, as well as their semimajor axes 
could be also inferred \citep[][]{bon12}.

In the case of \textit{binary 2}, which is more realistic due to smaller mass 
ratio $q$, the ripple effects in the composite line profiles are much weaker, 
and hence it is more difficult to resolve the constituent profiles and measure 
their Doppler shifts (see Fig. \ref{fig4}). The weak ripples occur mostly in 
the line core (see the first three panels of Fig. \ref{fig4}), except when the 
constituent line originating from the primary disk is much narrower than the 
corresponding profile emitted from the secondary disk (see the last panel of 
Fig. \ref{fig4}). In the latter case, both wings of the composite profiles 
(especially the ''blue'' one) could be also affected  by formation of the 
''bump''-like structures which vary due to Doppler shifts. In the 
former case of variability in the Fe K$\alpha$ line core, one should keep 
in mind that the relatively narrow line core in the case of AGN could have 
multiple origins which, besides an accretion disk, also include a parsec-scale 
molecular torus and, to a lesser extent, the optical broad-line region 
\citep{nan06}. Therefore, the mixing of the broad emission from the accretion 
disk with the narrow emission from the other regions could limit the 
possibility to detect the ripple effects due to Doppler shifts caused by 
orbital motion of the SMBH binary.

However, even in such cases when the ripple effects are weak, it could be 
possible to study the amplitudes and periods of their oscillations through the 
analysis of the light curves of different parts of the composite Fe K$\alpha$ 
lines, assuming that these oscillations are consequence of the binary SMBHs and 
not of some other phenomena, such as orbiting bright spots in the disk 
\citep[see e.g.][]{iwa04,jov10}. For instance, a recent reverberation study of 
the Fe K$\alpha$ light curves by \citet{zog13} for several different AGN 
demonstrated the feasibility of these type of investigations. As a result, the 
constraints to the mass ratios $q$ and orbital periods $P$ could be obtained, 
provided that the observed X-ray spectra of the studied SMBH binaries have 
sufficient signal to noise ratio.

As it can be seen from Figs. \ref{fig3} and \ref{fig4}, two assumed 
models of the accretion disks around the SMBHs in a binary system resulted with 
a variety of the composite Fe K$\alpha$ line profiles. However, one should keep 
in mind that for different parameters of two accretion disks (e.g. for 
different boundaries of the Fe K$\alpha$ line emitting region), the resulting 
composite line profiles could be even more diverse.

\section{Conclusions}

In this paper we simulated the composite profiles of the Fe K$\alpha$ spectral 
line emitted from two relativistic accretion disks in a binary system of 
SMBHs during its different orbital phases, in order to find whether such 
profiles and their variations could be detected and used for studying the 
properties of the SMBH binaries. From our investigations we can outline the 
following conclusions:

\begin{enumerate}
\item
The performed simulations showed that the composite Fe K$\alpha$ lines 
with rippled profiles could provide evidence about the presence of the binary 
SMBH systems, and could be used for studying their properties and orbits;
\item The most favorable candidates for such studies are the binaries with 
high mass ratios and radial velocities which reach $\gtrsim 10\%$ of the 
constituent line FWHMs; 
\item Mass ratios of the components and inclinations of their 
accretion disks have significant impact on detected composite Fe K$\alpha$ line 
variability caused by Doppler shifts due to orbital motion;
\item Such variability is in the form of ripple effects which oscillate with 
orbital period;
\item In the case of SMBH binaries with mass ratios approaching to 1, these 
ripple effects are very strong and the corresponding constituent profiles, as 
well as their Doppler shifts, could be resolved;
\item Besides the variability of the composite Fe K$\alpha$ line profiles, the 
properties of the observed SMBH binaries could be also constrained by studying 
the light curves of the different parts of the composite line, and in some 
cases, the radial velocity curves of the constituent line profiles originating 
from the primary and secondary accretion disks.

\end{enumerate}


\begin{thebibliography}{}


\bibitem[Amaro-Seoane et al.(2013)]{ama13} Amaro-Seoane, P., Aoudia, S., 
Babak, S., et al., eLISA/NGO: Astrophysics and cosmology in the 
gravitational-wave millihertz regime, GW Notes Special Issue: eLISA/NGO, 
revealing a hidden universe, 6, 4--110, 2013.

\bibitem[Artymowicz \& Lubow(1994)]{art94} Artymowicz, P., \& Lubow, 
S.~H., Dynamics of binary-disk interaction. 1: Resonances and disk gap sizes, 
ApJ, 421, 651--667, 1994.

\bibitem[Bogdanovi{\'c} et al.(2011)]{bog11} Bogdanovi{\'c}, T., Bode, T., Haas, 
R., Laguna, P., \& Shoemaker, D., Properties of accretion flows around 
coalescing supermassive black holes, Classical and Quantum Gravity, 28, 
id. 094020, 1--13, 2011. 

\bibitem[Bogdanovi{\'c} et al.(2009a)]{bog09a} Bogdanovi{\'c}, T., Eracleous, 
M., \& Sigurdsson, S., Emission lines as a tool in search for supermassive 
black hole binaries and recoiling black holes, New Astronomy Reviews, 53, 
113--120, 2009a. 

\bibitem[Bogdanovi{\'c} et al.(2009b)]{bog09b} Bogdanovi{\'c}, T., Eracleous, 
M., \& Sigurdsson, S., SDSS J092712.65+294344.0: Recoiling Black Hole or a 
Subparsec Binary Candidate?, ApJ, 697, 288--292, 2009b.

\bibitem[Bogdanovi{\'c} et al.(2008)]{bog08} Bogdanovi{\'c}, T., Smith, B.~D., 
Sigurdsson, S., \& Eracleous, M., Modeling of Emission Signatures of 
Massive Black Hole Binaries. I. Methods, ApJS, 174, 455--480, 2008. 

\bibitem[Bon et al.(2012)]{bon12} Bon, E., Jovanovi{\'c}, P., Marziani, P., et 
al., The First Spectroscopically Resolved Sub-parsec Orbit of a Supermassive 
Binary Black Hole, ApJ, 759, article id. 118, 1--8, 2012. 

\bibitem[Braxmaier et al.(2012)]{bra12} Braxmaier, C., Dittus, H., 
Foulon, B., et al., Astrodynamical Space Test of Relativity using Optical 
Devices I (ASTROD I)--a class-M fundamental physics mission proposal for cosmic 
vision 2015-2025: 2010 Update, Experimental Astronomy, 34, 181--201, 2012.

\bibitem[\v Cade\v z et al.(1998)]{cad98} \v Cade\v z, A., Fanton,
C., \& Calvani, M., Line emission from accretion discs around black holes: the 
analytic approach, New Astronomy, 3, 647--654, 1998.

\bibitem[De Paolis et al.(2003)]{dep03} De Paolis, F., Ingrosso, G., 
Nucita, A.~A., \& Zakharov, A.~F., Binary black holes in Mkns as sources of 
gravitational radiation for space based interferometers, A\&A, 410, 741--747, 
2003.

\bibitem[Fabian \& Ross(2010)]{fab10} Fabian, A. C., \& Ross, R. R., 
X-ray Reflection, Space Science Reviews, 157, 167--176, 2010.

\bibitem[Fabian et al.(1989)]{fab89} Fabian, A. C., Rees, M. J., 
Stella, L., \& White, N. E., X-ray fluorescence from the inner disc in Cygnus 
X-1, MNRAS, 238, 729--736, 1989.

\bibitem[Fanton et al.(1997)]{fan97} Fanton, C., Calvani, M.,
Felice, F., \& \v Cade\v z, A., Detecting Accretion Disks in Active Galactic 
Nuclei, Publications of the Astronomical
Society of Japan, 49, 159--169, 1997.

\bibitem[Gould(1995)]{gou95} Gould, A. Self-lensing by Binaries, ApJ, 446, 
541--542, 1995.

\bibitem[Hayasaki(2011)]{hay11} Hayasaki, K., Radiatively Inefficient 
Accretion Flows Induced by Gravitational-wave Emission Before Massive Black 
Hole Coalescence, ApJ, 726, article id. L14, 1--5, 2011.

\bibitem[Hilditch(2001)]{hild01} Hilditch, R. W., An Introduction
to Close Binary Stars, Cambridge University Press (New York, 2001).

\bibitem[Iwasawa et al.(2004)]{iwa04} Iwasawa, K., Miniutti, G., \& Fabian, 
A.~C., Flux and energy modulation of redshifted iron emission in NGC 3516: 
implications for the black hole mass, MNRAS, 355, 1073--1079, 2004.

\bibitem[Jovanovi{\'c}(2012)]{jov12} Jovanovi{\'c}, P., The broad Fe K$\alpha$ 
line and supermassive black holes, New Astronomy Reviews, 56, 37--48, 2012.

\bibitem[Jovanovi{\'c} et al.(2011)]{jov11} Jovanovi{\'c}, P., Borka 
Jovanovi{\'c}, V., \& Borka, D., Influence of Black Hole Spin on the Shape of 
the Fe K$\alpha$ Spectral Line: the Case of 3C 405, Baltic Astronomy, 20, 
468--471, 2011.

\bibitem[Jovanovi{\'c} \& Popovi{\'c}(2008)]{jov08a} Jovanovi{\'c}, P., \& 
Popovi{\'c}, L.~{\v C}., Observational effects of strong gravity in vicinity of 
supermassive black holes, Fortschritte der Physik, 56, 456--461, 2008.

\bibitem[Jovanovi{\'c} \& Popovi{\'c}(2009)]{jov09b} Jovanovi{\'c},
P. \& Popovi{\'c}, L.~{\v C}., X-ray Emission From Accretion Disks of AGN: 
Signatures of Supermassive Black Holes, in ''Black Holes and Galaxy
Formation'', eds. Adonis D. Wachter and Raphael J. Propst, Nova
Science Publishers Inc, Hauppauge NY, USA, 249--294, 2009, arXiv:0903.0978.

\bibitem[Jovanovi{\'c} et al.(2009)]{jov09a} Jovanovi{\'c}, P.,
Popovi{\'c}, L.~{\v C}., \& Simi{\'c}, S., Influence of gravitational 
microlensing on broad absorption lines of QSOs: The case of the Fe K$\alpha$ 
line, New Astronomy Reviews, 53, 156--161, 2009. 

\bibitem[Jovanovi{\'c} et al.(2010)]{jov10} Jovanovi{\'c}, P., Popovi{\'c}, 
L.~{\v C}., Stalevski, M., \& Shapovalova, A.~I., Variability of the H$\beta$ 
Line Profiles as an Indicator of Orbiting Bright Spots in Accretion Disks of 
Quasars: A Case Study of 3C 390.3, ApJ, 718, 168--176, 2010. 

\bibitem[Jovanovi{\'c} et al.(2008)]{jov08b} Jovanovi{\'c}, P., Zakharov, 
A.~F., Popovi{\'c}, L.~{\v C}., \& Petrovi{\'c}, T., Microlensing of the X-ray, 
UV and optical emission regions of quasars: simulations of the time-scales and 
amplitude variations of microlensing events, MNRAS, 386, 397--406, 2008. 

\bibitem[Lin \& Papaloizou(1979a)]{lin79a} Lin, D. N. C., \& Papaloizou, 
J., Tidal torques on accretion discs in binary systems with extreme mass ratios, 
MNRAS, 186, 799--812, 1979a.

\bibitem[Lin \& Papaloizou(1979b)]{lin79b} Lin, D. N. C., \& Papaloizou, 
J., On the structure of circumbinary accretion disks and the tidal evolution of 
commensurable satellites, MNRAS, 188, 191--201, 1979b.

\bibitem[McKernan et al.(2013)]{mck13} McKernan, B., Ford, K.~E.~S., Kocsis, B., 
\& Haiman, Z., Ripple effects and oscillations in the broad Fe K$\alpha$ line as 
a probe of massive black hole mergers, MNRAS, 432, 1468--1482, 2013. 

\bibitem[Nandra(2006)]{nan06} Nandra, K., On the origin of the iron 
K$\alpha$ line cores in active galactic nuclei, MNRAS, 368, L62--L66, 2006.

\bibitem[Nandra et al.(1997)]{nan97} Nandra, K., George, I.~M., Mushotzky, 
R.~F., Turner, T.~J., \& Yaqoob, T., ASCA Observations of Seyfert 1 Galaxies. 
II. Relativistic Iron K$\alpha$ Emission, ApJ, 477, 602--622, 1997.

\bibitem[Nandra et al.(2007)]{nan07} Nandra, K., O'Neill, P.~M., George, I.~M., 
\& Reeves, J.~N., An XMM-Newton survey of broad iron lines in Seyfert galaxies, 
MNRAS, 382, 194--228, 2007.

\bibitem[Popovi{\'c}(2012)]{pop12b} Popovi{\'c}, L.~{\v C}., Super-massive 
binary black holes and emission lines in active galactic nuclei, New Astronomy 
Reviews, 56, 74--91, 2012.

\bibitem[Popovi{\'c} et al.(2003a)]{pop03a} Popovi{\'c}, L.~{\v C}., 
Jovanovi{\'c}, P., Mediavilla, E., \& Mu{\~n}oz, J.~A., Influence of 
Microlensing on the Active Galactic Nucleus Fe Kalpha Line, Astronomical
and Astrophysical Transactions, 22, 719--725, 2003a. 

\bibitem[Popovi{\'c} et al.(2006)]{pop06} Popovi{\'c}, L.~{\v C}., 
Jovanovi{\'c}, P., Mediavilla, E., et al., A Study of the Correlation between 
the Amplification of the Fe K$\alpha$ Line and the X-Ray Continuum of Quasars 
due to Microlensing, ApJ, 637, 620--630, 2006. 

\bibitem[Popovi{\'c} et al.(2005)]{pop05} Popovi{\'c}, L.~{\v C}., 
Jovanovi{\'c}, P., Petrovi{\'c}, T., \& Shalyapin, V.~N., Amplification and 
variability of the AGN X-ray emission due to microlensing, Astronomische 
Nachrichten, 326, 981--984, 2005.

\bibitem[Popovi{\'c} et al.(2012)]{pop12a} Popovi{\'c}, L.~{\v C}., 
Jovanovi{\'c}, P., Stalevski, M., et al., Photocentric variability of quasars 
caused by variations in their inner structure: consequences for Gaia 
measurements, A\&A, 538, article id. A107, 1--11, 2012. 

\bibitem[Popovi{\'c} et al.(2003b)]{pop03b} Popovi{\'c}, L.~{\v C}., 
Mediavilla, E.~G., Jovanovi{\'c}, P., \& Mu{\~n}oz, J.~A., The influence of 
microlensing on the shape of the AGN Fe K$\alpha$ line, A\&A, 398, 975--982, 
2003b. 

\bibitem[Reynolds \& Nowak(2003)]{rey03} Reynolds, C. S., \& Nowak, M. 
A., Fluorescent iron lines as a probe of astrophysical black hole systems, 
Physics Reports, 377, 389--466, 2003.

\bibitem[Sesana et al.(2012)]{ses12} Sesana, A., Roedig, C., Reynolds, M. T., 
\& Dotti, M., Multimessenger astronomy with pulsar timing and X-ray observations 
of massive black hole binaries, MNRAS, 420, 860--877. 2012.

\bibitem[Shen et al.(2013)]{she13} Shen, Y., Liu, X., Loeb, A., \& Tremaine, 
S., Constraining Sub-parsec Binary Supermassive Black Holes in Quasars with 
Multi-epoch Spectroscopy. I. The General Quasar Population, ApJ, 775, article 
id. 49, 1--23, 2013.

\bibitem[Tanaka et al.(1995)]{tan95} Tanaka, Y., Nandra, K., Fabian, A. 
C., Inoue, H., Otani, C., Dotani, T., Hayashida, K., Iwasawa, K., Kii, T., 
Kunieda, H., Makino, F., Matsuoka, M., Gravitationally redshifted emission 
implying an accretion disk and massive black hole in the active galaxy 
MCG-6-30-15, Nature, 375, 659--661, 1995.

\bibitem[Yu \& Lu(2001)]{yul01} Yu, Q., \& Lu, Y., A\&A, Fe K$\alpha$ 
line: A tool to probe massive binary black holes in Active Galactic Nuclei?, 
377, 17--22, 2001.

\bibitem[Zoghbi et al.(2013)]{zog13} Zoghbi, A., Reynolds, C., Cackett, E.~M., 
Miniutti, G., Kara, E., Fabian, A. C., Discovery of Fe K$\alpha$ X-Ray 
Reverberation around the Black Holes in MCG-5-23-16 and NGC 7314, ApJ, 767, 
article id. 121, 1--6, 2013.

\end{thebibliography}
\end{document}